# Scintillation properties of CsPbBr$_3$ nanocrystals prepared by ligand-assisted re-precipitation and dual effect of polyacrylate encapsulation towards scalable ultrafast radiation detectors

Francesca Cova[1†], Andrea Erroi[1†], Matteo L. Zaffalon[1], Alessia Cemmi[2], Ilaria Di Sarcina[2], Jacopo Perego[1], Angelo Monguzzi[1], Angiolina Comotti[1], Francesca Rossi[3], Francesco Carulli[1*], and Sergio Brovelli[1*]

[1] Department of Materials Science, University of Milano - Bicocca, via Roberto Cozzi 55, 20125 Milano, Italy
[2] ENEA Fusion and Technology for Nuclear Safety and Security Department, via Anguillarese 301, 00123 Casaccia R.C. Rome, Italy
[3] IMEM-CNR Institute, Parco Area delle Scienze, 37/A, 43124, Parma, Italy

*sergio.brovelli@unimib.it, francesco.carulli@unimib.it

**ABSTRACT**

**Lead halide perovskite nanocrystals (LHP-NCs) embedded in polymeric hosts are gaining attention as scalable and low-cost scintillation detectors for technologically relevant applications. Despite rapid progress, little is currently known about the scintillation properties and stability of LHP-NCs prepared by the Ligand Assisted Re-Precipitation (LARP) method, which allows mass scalability at room temperature unmatched by any other type of nanostructure, and the implications of** 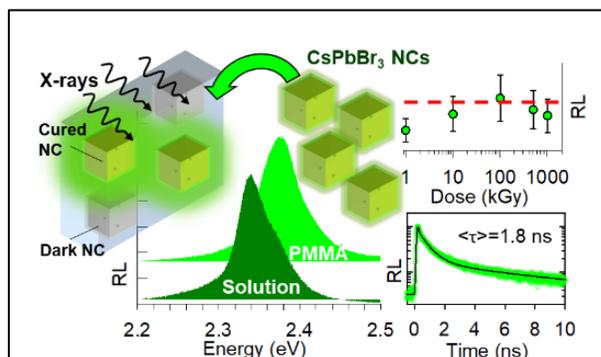 **incorporating LHP-NCs into polyacrylate hosts are still largely debated. Here, we show that LARP-synthesised CsPbBr$_3$ NCs are comparable to particles from hot-injection routes and unravel the dual effect of polyacrylate incorporation, where the partial degradation of LHP-NCs luminescence is counterbalanced by the passivation of electron-poor defects by the host acrylic groups. Experiments on NCs with tailored surface defects show that the balance between such antithetical effects of polymer embedding is determined by the surface defect density of the NCs and provide guidelines for further material optimisation.**

The detection of ionizing radiation is critical in many applications including nuclear and homeland security[1, 2], high-energy physics[3, 4], and medicine[5, 6]. To date, massive inorganic scintillator crystals account for the largest share of the market but suffer from high production costs and limited customization[7, 8]. Plastic scintillators, on the other hand, have faster kinetics and can be fabricated in custom size and shape at low cost, but suffer from poor radiation stability and lower detection efficiency. In an attempt to combine the advantages of these two classes of scintillators, increasing attention is being paid to the development of nanocomposite scintillators consisting of polymeric matrices doped or coated with nanocrystals (NCs) of high atomic number compounds prepared by scalable solution-based processes[9-13]. Metal halide NCs, and in particular lead halide perovskites (LHP), have emerged as promising scintillators[10, 14-21], prized for their solution synthesis, efficient color-tunable radioluminescence and high radiation hardness[22]. To date, most studies have been devoted to understanding and optimizing the scintillation of LHP-NCs prepared by the hot-injection method[20, 23], which produces the most luminescent materials. Specifically, recent studies of $CsPbBr_3$ NCs have shown that the scintillation emission is largely multi-excitonic[19] with substantial contributions by shallow trapped excitons in surface bromine vacancies[20, 23, 24], which can be suppressed by various resurfacing techniques[25-29], resulting in enhanced light yield and exceptionally high radiation stability[20, 30]. Despite such advantages, hot-injection synthesis is performed at high temperature and is hardly scalable to the very large quantities required for radiation detection. On the other hand, LHP-NCs can be produced at room temperature by the so-called ligand-assisted reprecipitation method (commonly referred to as LARP)[31-37]. Because of their industrial relevance, LARP-synthesized LHP-NCs have received considerable attention leading to the realization of effective strategies to suppress non-radiative defects and increase the luminescence quantum yield ($\Phi_{PL}$) up to 90%[38-40]. Despite this progress, the study of LARP-synthesized LHP-NCs as nanoscintillators still lags behind, and the scintillation process and the competing non-radiative (mostly trap-related) phenomena remain largely unexplored, as does their radiation stability both as bare NCs and in nanocomposites. Indeed, another often largely overlooked critical aspect in the fabrication of nanocomposite scintillators is the effect of embedding on the scintillation properties and competing charge/exciton trapping mechanisms[41]. Specifically, optical spectroscopy studies agree that the embedding in non-polar polymers, such as polystyrene (PS), leaves the optical properties of LHP-NCs largely unaltered and enhances their stability against oxygen and moisture through a sealing effect[42, 43]. Unfortunately, the mass polymerization of PS and its derivatives requires thermal polymerization approaches at high temperature (≥80°C)[44-46] that typically degrade the NCs and/or cause phase transitions to non-emissive allotropes[47, 48]. On the other hand, polyacrylates, such as polymethyl methacrylate (PMMA), can be polymerized with less aggressive photoinitiated routes but their effect on LHP-NCs is still debated[49, 50]. Specifically, studies showed the deterioration of the NC surfaces due to the partial polarity of the acrylic moieties, leading to a lower emission efficiency[10, 51], while others demonstrated the beneficial surface passivation of undercoordinated cation sites by the carboxylate-methacrylate units, leading to increased luminescence yield or photovoltaic performance[52-58]. The disambiguation of this dual effect of polyacrylates is particularly relevant for LHP-NCs based plastic scintillators because, under ionizing radiation excitation, defects in the NCs surfaces and/or at the NC/polymer interface could act as traps for the NCs excitons and for free charge carriers, resulting in low light yield, slow scintillation tails and delayed recombination pathways detrimental to the scintillation performance[59, 60].

In this work, we aim to contribute to the advancement of this field by investigating the scintillation properties of $CsPbBr_3$ NCs synthesized via LARP and the effect of their incorporation into PMMA nanocomposites. Our results fill a gap in the knowledge of the scintillation physics of LHP NCs and clarify the controversial role of the polyacrylate host in the scintillation properties.

$CsPbBr_3$ NCs were synthesized according to the LARP protocol from ref.61. Consistent with previous reports[31, 40], high-resolution transmission electron microscopy (HRTEM) shows the formation of a mixture of particles with different aspect ratio, with a predominance of cubic NCs with average size of 14±3 nm (Figure S1). To produce self-standing PMMA nanocomposites, $CsPbBr_3$ NCs (0.5wt% compared to polymer) were dispersed in a 120 mg PMMA solution, and the solvent was slowly evaporated over 22 hours following the procedure described in ref.10 (see Methods for details). For the purpose of this study, we specifically opted for this fabrication procedure over mass polymerization in order to avoid potential parasitic effects by the radical initiators[62]. Photographs of $CsPbBr_3$ NCs in a toluene solution and in a PMMA nanocomposite under ambient and UV light are shown in **Figure 1a**. The X-ray diffraction (XRD) patterns of the NCs powder and nanocomposite (Figure S1) show main diffraction peaks at 21.5°, 30.7° and 37.7° matching the orthorhombic phase of $CsPbBr_3$, characterized by the typical splitting of the two latter peaks, while the broad diffraction peak at 13.7° of the nanocomposite film and is attributed to the contribution of the PMMA[63, 64]. The optical absorption and PL spectra of $CsPbBr_3$ NCs in solution and in PMMA (**Figure 1b**) show identical absorption edge at 2.52 eV, and the PL peaks at 2.44 eV in both cases, indicating that the electronic structure and particle size of the NCs are preserved upon embedding. Nonetheless, polymer encapsulation has substantial effects on the optical properties of the NCs: on the one hand, the PL efficiency of the NCs ($\Phi_{PL}$=34±5%) decreases to $\Phi_{PL}$=17±5% in the nanocomposite due to the degradation of LHP-NCs in polar hosts[38, 54, 55, 65-69]. On the other hand, the inspection of the PL kinetics (**Figure 1c**) reveals the typical signatures of reduced trapping losses. Specifically, as previously reported, the PL kinetics can be tentatively modelled with a triple exponential function with a fast component (with lifetime, $\tau_F$) due to band-edge excitons affected by trapping and an intermediate one due to mostly radiative recombination of band edge excitons ($\tau_X$) followed by delayed band-edge luminescence ($\tau_S$) ascribed to thermal release of excitons trapped in shallow defect states[20, 23, 70, 71]. Incorporation into PMMA slows down the fast component (Supporting Table 1) and reduces the weight of the delayed PL from 50% to 25%. Spectrally resolved PL decay (**Figure 1d**) shows no change in the PL with time in both cases, confirming that all the decay components are related to band-edge luminescence. The data therefore point to a dual effect of PMMA, causing the ultrafast 'static' quenching of a sub-fraction of the NCs ensemble that occurs prior to radiative decay (and therefore does not cause an acceleration of the emission kinetics), and on the other hand an enhanced surface passivation of the remaining NCs fraction[53, 72], whose emission is slower and purged of contributions from trapped exciton recombination (sketch in **Figure 1e**).

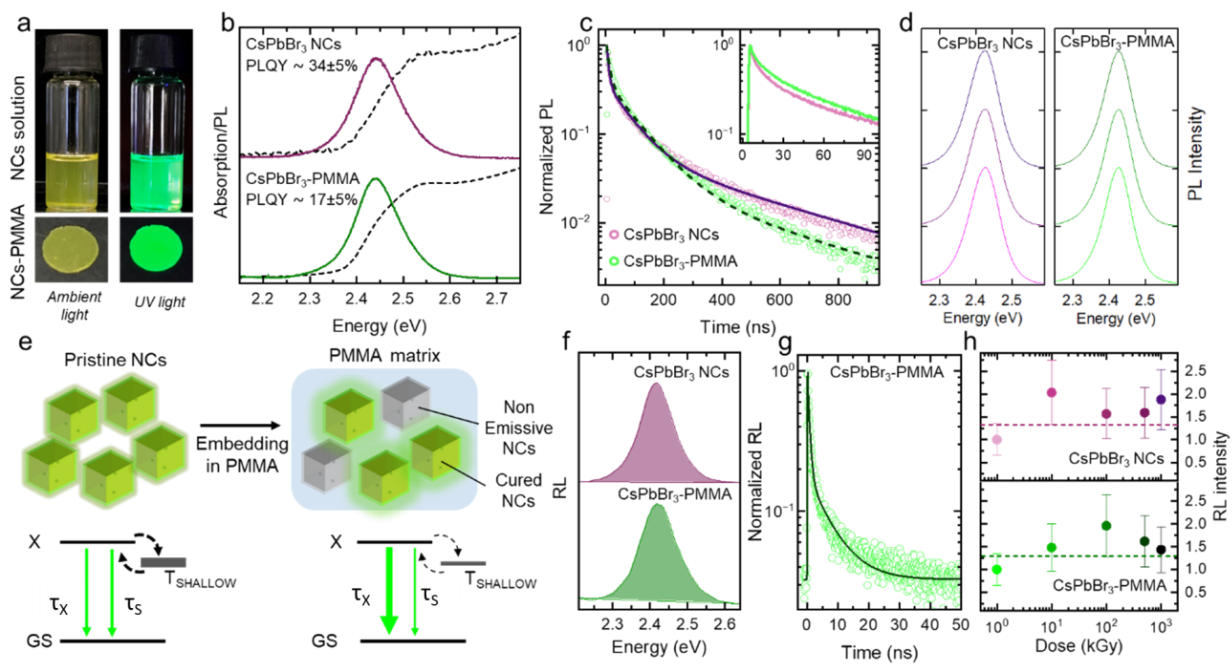

**Figure 1.** a) Photographs of a toluene solution of $CsPbBr_3$ NCs and of a PMMA nanocomposite film under ambient and UV light. b) Optical absorption (dashed black lines) and PL (solid line) spectra of $CsPbBr_3$ NCs in toluene solution and nanocomposite. c) Normalized PL decay curves (open symbols) measured at the PL maximum by exciting with 3.05 eV light $CsPbBr_3$ NCs in toluene solution and nanocomposite. Fitting curves are reported as lines. d) PL spectra of a toluene solution of $CsPbBr_3$ NCs and of a PMMA nanocomposite film obtained integrating their respective PL decay traces in 0-20 ns (bottom), 40-150 ns (middle) and 200-950 ns (top) time range. e) Schematic representation of the effects of embedding in PMMA on the excitonic recombination of LHP-NCs. f) RL spectra of $CsPbBr_3$ NCs and of a PMMA nanocomposite. g) RL decay trace (open symbols) for the PMMA nanocomposite measured at the RL maximum. Fitting curve is reported as solid line. h) RL intensity of $CsPbBr_3$ NCs and PMMA nanocomposite as a function of cumulative irradiation dose up to 1 MGy. Dashed lines are guides for the eyes.

As counterevidence, NCs embedded in PS, which has no acrylic functionalities, show PL dynamics essentially identical to those of NCs in solution (Figure S2). Consistent with previous results, the room temperature RL spectra (**Figure 1f**) of both the NCs and the nanocomposite are slightly red-shifted with respect to the corresponding PL, owing to the contribution of attractive biexcitons[19] and trapped excitons in bromine surface vacancies[73] at the NCs surfaces that are likely preferentially populated by electrons resulting from the thermalization of highly energetic carriers generated upon the primary interaction with X-rays[20]. Further compelling evidence for the substantial passivation effect of the PMMA embedding of NCs surface defects comes from low temperature PL and RL spectra reported below showing the suppression of shallow defect RL contribution for the nanocomposite. Before proceeding with such a detailed analysis, we performed time-resolved scintillation and gamma irradiation experiments. The time-resolved RL of the CsPbBr$_3$ NCs in PMMA are reported in **Figure 1g** showing, in agreement with recent reports[11, 19, 74], a largely ultrafast kinetics (average lifetime $<\tau>$=1.81 ns) with a prompt decay – limited by the ~75 ps time response of our setup – and a ~600 ps contribution respectively ascribed to the recombination of multi- and charged-excitons formed under ionizing radiation, followed by a ~7 ns tail matching the fast PL component of neutral band-edge excitons (see Supporting Information for the fitting procedure). Such a fast scintillation kinetics confirms the potential of LARP-synthesized LHP-NCs for fast timing applications further corroborated by the mass scalability of the LARP method[32, 39]. Next, in order to assess the radiation stability of our systems, we used the Calliope irradiation facility[75] to expose both the NCs and the nanocomposite samples to uniform γ-ray irradiation from a $^{60}$Co source at a dose rate of 3.05 kGy$_{air}$ h$^{-1}$ and monitored their optical and scintillation properties at increasing cumulative doses up to 1 MGy. For the tested samples containing 0.5 wt % NCs, the mass and linear attenuation coefficients are μ/ρ = 6.185×10$^{-2}$ cm$^2$ g$^{-1}$ and μ = 7.008×10$^{-2}$ cm$^{-1}$, respectively, resulting in a cumulative dose of 2.3 MGy. Remarkably, the RL of both the NCs and the nanocomposite is stable over 6 orders of magnitude of the delivered dose (**Figure 1h**), consistent with the recently demonstrated exceptional radiation hardness of CsPbBr$_3$ NCs[19, 20]. We stress that radiation hardness studies using $^{60}$Co on various types of plastic scintillators showed massive light yield drops (40-80% decrease) upon gamma doses of only 100 kGy[76]. Consistently, we observed the rise of high-energy radiation-induced absorption band in bare PMMA exposed to identical irradiation conditions as the NCs and nanocomposites (Figure S3). The stability of the optical and scintillation properties of the nanocomposite has been further assessed also under prolonged X-ray irradiation and UV light excitation producing comparable exciton occupancy (Figure S4). Taken together, the results shown in **Figure 1** indicate that the scintillation behavior, timing and stability of CsPbBr$_3$ NCs prepared by the LARP method are comparable to those of NCs from hot-injection synthesis (except for the typically observed lower $\Phi_{PL}$ value in the absence of surface post-treatments)[20], which is promising for mass scaling purposes, and that polymer embedding could be exploited as a possible strategy to reduce the defect density at the particle surfaces.

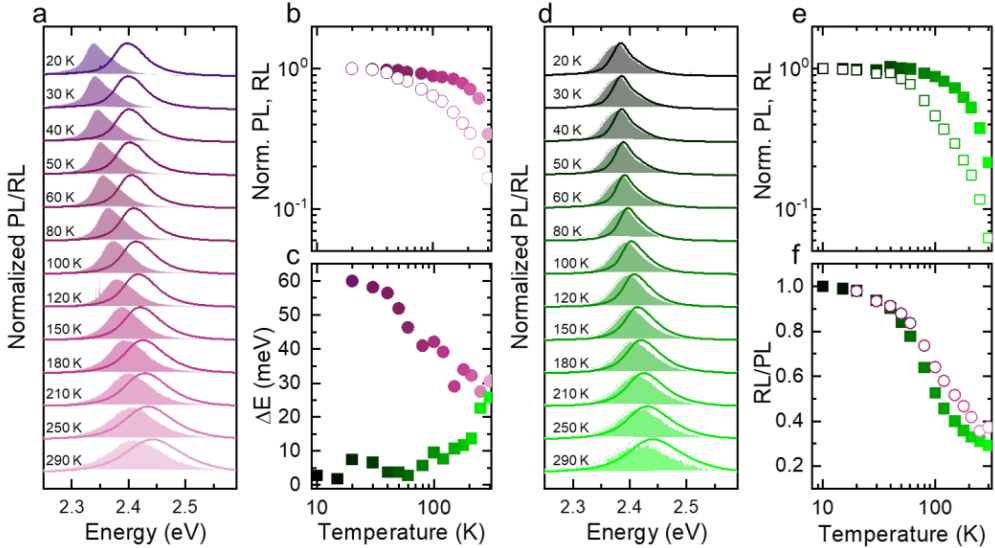

**Figure 2.** a) PL (solid line) and RL (shaded area) spectra of CsPbBr$_3$ NCs spin-casted onto an Al$_2$O$_3$ substrate (a) as a function of temperature from 290 to 20 K (from bottom to top as indicated in figure). b) Spectrally integrated total PL (filled symbols) and RL (hollow symbols) of CsPbBr$_3$ NCs as a function of temperature. c) Shift in energy between the PL and RL peak as a function of temperature for the CsPbBr$_3$ NCs (purple markers) and nanocomposite (green markers). d) Same as (a) for PMMA nanocomposite. e) Same as (b) for PMMA nanocomposite. f) RL intensity normalized to the PL intensity variation with temperature for the CsPbBr$_3$ NCs (purple markers) and nanocomposite (green markers).

To gain deeper insights into the role of defects and polymer embedding in the scintillation of LARP-synthesized CsPbBr$_3$ NCs, we performed PL and RL measurements at controlled temperature down to T=20 K, where thermal release of trapped carriers is largely inhibited[77]. Upon cooling, the PL spectra of bare CsPbBr$_3$ NCs gradually redshift and narrow (**Figure 2a**) due to lattice expansion and reduced phonon coupling[78]. At the same time, $\Phi_{PL}$ increases by a factor of three, suggesting that at 20 K the PL process is almost completely radiative (**Figure 2b**). Consistent with previous results on CsPbBr$_3$ NCs from hot-injection synthesis, the RL spectrum of the bare NCs shows a gradually stronger contribution from shallowly trapped excitons upon cooling, which is suggested by the increasingly larger redshift from the corresponding PL and its increasing multicomponent nature, ascribed to a distribution of shallow emissive defects[20, 23] (**Figure 2c**). The RL intensity shows stronger intensification (**Figure 2b**), as further emphasized by the temperature evolution of the $I_{RL}/I_{PL}$ ratio shown in **Figure 2f**, suggesting that the RL process is influenced by other non-radiative pathways such as trapping of diffusely generated free carriers in deep trap sites (also in line with the observed redshift of the RL vs PL spectra discussed above), that are also suppressed at cryogenic temperature. Moreover, the evidence of shallow trapped excitons at low temperature indicates that the optical properties improvement upon polymer encapsulation are not due to restricted degrees of freedom of the surface ligand shell causing nonradiative losses, as these would be also suppressed at T<50 K. Crucially, the incorporation of NCs into PMMA restores the band-edge origin of RL (**Figure 2d**) which now matches the corresponding PL except for the visible low energy shoulder, ascribed to the biexcitonic contribution to the scintillation emission[11, 19], and shows a more pronounced enhancement with temperature decreasing (**Figure 2e**). These observations support the partial passivation of surface defects due to the coordination of the electron-rich acrylic groups of PMMA with uncoordinated Pb sites on the NCs surfaces[54, 79].

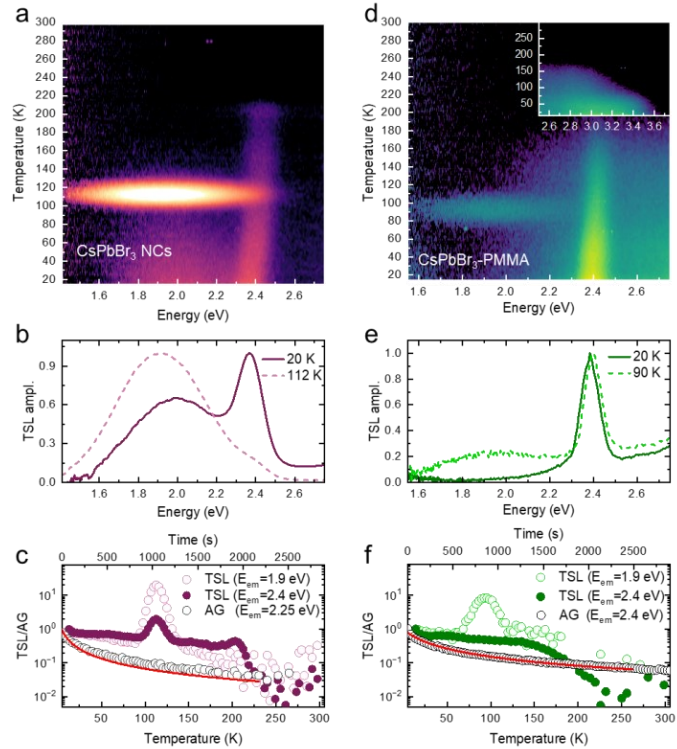

**Figure 3.** a) Contour plot of the spectrally resolved TSL as a function of temperature for CsPbBr$_3$ NCs after X-ray irradiation at 10 K. b) Normalized TSL spectra at different temperatures (reported in the legend) and c) glow curve at different emission energies (E$_{em}$) extracted from (a) and AG decay at 10 K (black open symbols). The red solid line is the fit of the AG decay to a power-law function described along the text. d) Same as (a) for the nanocomposite film. Inset: contour plot of the TSL amplitude in the high energy region, corresponding to emission of the PMMA matrix. e, f) Same as (b,c) for the nanocomposite.

To extend our investigation of the role of deep trap states and possible effects of polymer embedding in the scintillation process of LARP-synthesized CsPbBr$_3$ NCs, we performed TSL and low-temperature AG experiments. The TSL contour plot of CsPbBr$_3$ NCs and the associated spectral and thermal analysis in **Figure 3a-c** show two distinct TSL features. The dominant emission at T<80 K peaks at 2.4 eV with a broad band extending in the low energy region, whereas a broad intragap emission at 1.9 eV appears at *T*=112 K (purple open symbols, **Figure 3c**), which is attributed to the complete depletion of deeper traps to an emissive defect by thermally assisted processes[80, 81]. This observation is consistent with previous reports on CsPbBr$_3$ single crystals[23] and CsPbBr$_3$ NCs treated by post-synthesis fluorination, where the same TSL signal was attributed to deep-state depletion responsible for weak slow scintillation tails[20]. Therefore, the appearance of this TSL peak in our LARP-synthesized NCs is ascribed to the presence of a deep trap state which becomes dominant likely because of the lower surface-to-volume ratio (S/V=0.42 nm$^{-1}$) compared to previously investigated hot-injection synthetized NCs (S/V=0.66 nm$^{-1}$)[20, 23]. The spectrally resolved AG profile, recorded at 10 K and plotted as a function of time in **Figure 3c** and Figure S5 closely resembles the low-*T* TSL spectral shape and follows a power-law function of type *I(t)=A(t+t$_0$)$^{-p}$*, where the value of p≈1 indicates that carrier release occurs from a distribution of non-radiative traps, in agreement with the broad TSL glow curve extending from 10 K to 200 K, evaluated at 2.4 eV (full symbols, **Figure 3c**). Such a trend is different compared to the behavior of hot-injection CsPbBr$_3$ NCs discussed in ref.20 where the depletion of shallow traps occurs via a-thermal tunneling among a dense defect network and might suggest that in this case the defects are either fewer in number or non-dispersed on the NCs surfaces. The TSL of the nanocomposite (**Figure 3d**) is similar to that of the bare NCs, but the 1.9 eV contribution is much weaker in intensity than the 2.4 eV emission, and the corresponding TSL peak occurs at significantly lower temperatures (**Figure 3f**, 90 K vs 112 K). Based on this, we tentatively attribute the slight change in the distribution of deep traps to the interaction with the acrylic groups of the polymer[58]. Indeed, a typical TSL measurement consists in a preliminary X-ray irradiation at low temperature to fill stable trap states: successively, the irradiation is ceased, and the temperature is progressively raised with a linear heating rate while monitoring the delayed emission due to carrier detrapping. The temperature increase makes traps unstable and provides the trapped carriers with the appropriate thermal energy to be freed and available for recombination. The emergence of TSL peaks in the glow curve is the signature of thermally activated detrapping processes since the rise of the TSL corresponds to the onset of carrier release from a trap and the TSL signal starts to decrease once a significant portion of the traps has been emptied. In an AG measurement, the delayed emission after the exposure to X-ray irradiation is monitored in isothermal conditions.

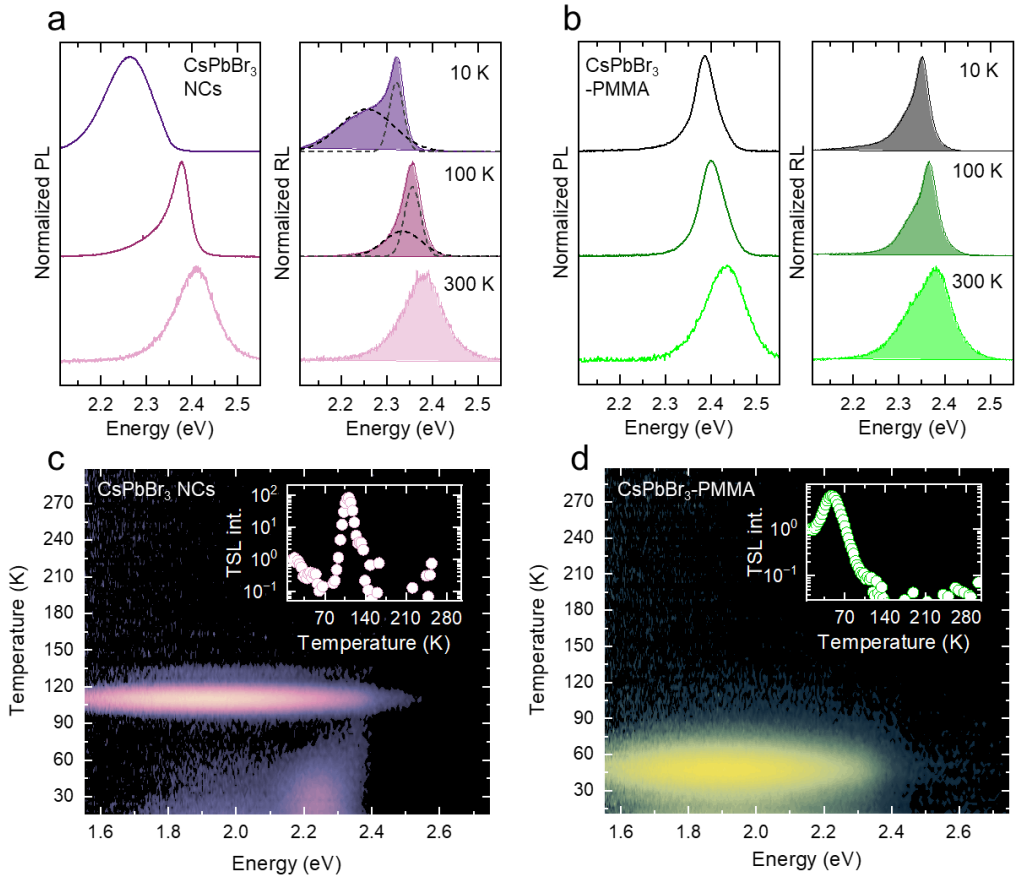

**Figure 4.** a) PL (left panel) and RL (right panel) spectra at 10 K, 100 K, and 300 K (from top to bottom) of defect-rich $CsPbBr_3$ NCs. Dashed black lines are the two Gaussian functions to emphasize the excitonic and the defect contributions to the RL spectrum. b) Same as 'a' for the PMMA nanocomposite. c) Contour plot of the spectrally resolved TSL for the defect-rich $CsPbBr_3$ NCs. Inset: glow curve at 1.9 eV, extracted from (c). d) Same as (c) for the corresponding PMMA nanocomposite.

To probe the effect of the surface defect density on the interplay between the beneficial and negative effects of PMMA, we prepared $CsPbBr_3$ NCs with high defect density (hereafter indicated as defect-rich NCs) following the modified LARP protocol reported in the Methods Section and incorporated them into PMMA nanocomposite using the same procedure described above. Interestingly, in this case, the emission efficiency increases dramatically from $\Phi_{PL}=2\pm1\%$ for the bare NCs to $20\pm5\%$ in the nanocomposite, confirming the strong surface passivation effect of PMMA. This is supported by the PL and RL measurements vs. *T* reported in **Figure 4a** and Figure S6: upon cooling, the PL (RL) spectrum of defect-rich NCs becomes largely dominated by a broad emission at ~2.25 eV due to nearly complete emission by trapped excitons. Incorporation of such NCs into PMMA restores the band-edge origin of PL even at low temperatures (**Figure 4b**). Side-by-side comparison between the TSL spectra of the bare NCs and the respective nanocomposite (**Figure 4c** and **4d**) reveal substantial differences. Specifically, the TSL contour plot of bare defect-rich NCs resembles that of the higher-quality counterparts in **Figure 3a**, with a low-$T$ emission at 2.25 eV due to shallow trapped excitons followed by a broad 1.9 eV luminescence at higher temperatures, indicating that the same traps and radiative defects are present in both systems with similar release mechanism. The ten-fold difference between the respective $\Phi_{PL}$-values strongly suggests that such deep traps are weakly involved in nonradiative emission losses but likely trap free carriers generated upon excitation with ionizing radiation. Notably, the nanocomposite containing defect-rich NCs shows only the emission band at 1.9 eV occurring at substantially lower temperature ($T$=50 K vs $T$=90 K) than for the higher quality counterparts. This indicates that in this case shallow defects are largely passivated and that the polymer plays an even stronger role in modifying the distribution of long-lived nonradiative traps responsible for its population at high defect density levels. To fully disambiguate the role of PMMA in the TSL of the nanocomposite, we performed the same TSL experiments on bare PMMA which returns a broad and featureless TSL profile typical of disordered materials related to the low-$T$ emission of the polymer extending up to 3.4 eV[56] (Figure S7).

In conclusion, we realized nanocomposite scintillators consisting of LARP-synthesized CsPbBr$_3$ NCs in a PMMA matrix and performed an in-depth spectroscopic investigation of the effect of polymer embedding on the scintillation properties and related (de)trapping phenomena. We also validated the ultrafast scintillation and high radiation resistance of plastic nanocomposites with LHP NCs up to extremely high γ-ray doses. The combined use of PL and RL measurements shed light on the thermal equilibrium between photogenerated excitons and local shallow states in the NCs, and TSL and afterglow experiments revealed the presence of deep trap states in both bare NCs and nanocomposites, releasing carriers mainly through thermally activated de-trapping pathways. Our results extend the knowledge of the scintillation properties of LHP-NCs to materials from highly scalable methods and clarify the role of polymer embedding in controlling the defects on NCs surfaces, which could be further exploited for the optimization of LHP-NCs-based nanocomposite scintillators.

**Associated Content**

**Supporting Information**

Detailed description of nanocrystal synthesis and nanocomposite fabrication, experimental methods and characterization data such as size distribution, TEM images and X-ray diffraction data. Optical absorption spectra of gamma irradiated samples and X-ray stability measurements. TSL spectra for bare PMMA matrix before and after gamma irradiation. Fitting procedures to extract scintillation kinetics parameters.

## Acknowledgements

This work was funded by the Horizon Europe EIC Pathfinder programme, project 101098649 - UNICORN and by the Italian Ministry of Research PRIN programme, project IRONSIDE.

# Supporting Information

Scintillation properties of CsPbBr$_3$ nanocrystals prepared by ligand-assisted re-precipitation and dual effect of polyacrylate encapsulation towards scalable ultrafast radiation detectors

Francesca Cova[1,†], Andrea Erroi[1,†], Matteo L. Zaffalon[1], Alessia Cemmi[2], Ilaria Di Sarcina[2], Jacopo Perego[1], Angelo Monguzzi[1], Angiolina Comotti[1], Francesca Rossi[3], Francesco Carulli[1,*], and Sergio Brovelli[1,*]

[1] *Department of Materials Science, University of Milano - Bicocca, via Roberto Cozzi 55, 20125 Milano, Italy*
[2] *ENEA Fusion and Technology for Nuclear Safety and Security Department, Casaccia R.C. Rome, Italy*
[3] *IMEM-CNR Institute, Parma, Italy*

*sergio.brovelli@unimib.it, francesco.carulli@unimib.it

**Experimental Section**

Materials

Oleic Acid (OA, 90%), Oleylamine (OAm, 90%), Lead bromide ($PbBr_2$, 98%), Cesium bromide (CsBr, 99.9%), Toluene (99%, anhydrous), N-N dimethylformamide (DMF, 99.8%), poly(methyl methacrylate) powder (PMMA, avg $M_W$ 15000), polystyrene rods (PS, avg $M_W$ 10000) were all purchased from Sigma Aldrich and were used directly without further purification.

Sample preparation

*Standard LARP synthesis of $CsPbBr_3$ NCs:* an optimized LARP protocol based on the work by Li et al. [1] was used as a $CsPbBr_3$ reference sample. Briefly, the precursor solution is prepared by dissolving 84.8 mg of CsBr and 146.8 mg of $PbBr_2$ in 10 mL of anhydrous DMF (0.040 M of CsBr, 0.040 M of $PbBr_2$). Subsequently 1 mL of OA, and 0.5 mL of OLA are added, and the solution is heated at 80 °C for 30 minutes until a clear, homogeneous solution is obtained. The LARP synthesis is then performed by quickly adding 1 mL of the precursor solution into a vigorously stirred anhydrous toluene solution at room temperature. Immediately after the injection, a quick swift of the solution color from transparent to bright green is noticed, confirming the correct formation of $CsPbBr_3$ NCs. The solution is kept under stirring for 20 minutes to ensure complete NCs formation. Finally, the $CsPbBr_3$ NCs are isolated through a multi-step centrifugation procedure without the use of any additional anti-solvent. Specifically, the solution is centrifuged at 8000 rpm for 10 min, the supernatant, containing small NCs, is discarded and the precipitate is collected and re-dissolved in 2 mL of clean toluene. Subsequently, the toluene solution is centrifuged at 3000 rpm for 5 minutes to remove any residual larger impurities: the precipitate (mainly composed by poorly capped NCs and unreacted reagents) is discarded, while the supernatant, containing the correctly formed NCs is taken for further measurements and stored under ambient conditions in the dark.

*Modified LARP synthesis of defect-rich $CsPbBr_3$ NCs*: in a 5 mL anhydrous DMF solution are added 42.6 mg of CsBr and 73.4 mg of $PbBr_2$ and the solution is kept under vigorous stirring. 0.5 mL of this solution is taken and 100 μL of OLA are added and kept under stirring at 60°C for 5 mins. LARP synthesis is performed by adding the whole OLA-DMF solution drop by drop into a vigorously stirred toluene solution (10 mL of total volume). The toluene solution progressively shifts from transparent to green during the addition of precursors solution. Finally, the isolation of $CsPbBr_3$ NCs was performed without the use of any additional antisolvent, centrifugating the crude solution at 3000 rpm for 5 mins to remove larger NCs and unreacted reagents and keeping the supernatant. Subsequently, the supernatant solution is centrifuged at 7000 rpm for 15 min and the precipitated is collected and re-dissolved in toluene to perform the optical characterization.

*Preparation of polymeric nanocomposites:* the nanocomposites were fabricated by evaporation of a PMMA or PS solution with dispersed $CsPbBr_3$ NCs. In a vial, 0.5 ml of a 1 mg ml$^{-1}$ solution of $CsPbBr_3$ NCs in toluene was added to 1 ml $CHCl_2$ containing 100 mg of PMMA or PS. On evaporation at 20 °C for 22 h under an almost saturated atmosphere of $CHCl_2$, smooth and homogeneous polymeric nanocomposites (0.3 mm thick, 12 mm diameter) were obtained.

Material characterization

*Structural characterization:* powder XRD experiments were performed on a Rigaku SmartLab SE diffractometer equipped with Cu Kα radiation, with a Cu Kβ radiation filter and a HyPix-400 detector, operating at 40 kV and 30 mA. Data collections were performed in Bragg–Brentano geometry with a 2θ scan range between 10.00° and 40.00° with a step size of 0.02° and scan speed of 0.10° min$^{-1}$. Powdery samples were gently dispersed on a silicon crystal sample holder (zero-background sample holder). NCs samples were drop-casted on a Carbon coated Cu grid for TEM analysis, in high-resolution mode and Z-contrast scanning mode, in a JEOL JEM-2200FS microscope operated at 200 kV.

*Optical measurements*: optical absorption spectroscopy was carried out using a Perkin Elmer Lambda 950 UV/VIS double beam spectrometer equipped with a Spectralon coated integrating sphere in order to collect scattered light. PL quantum yield ($\Phi_{PL}$) measurements were performed on the same solutions using an integrating sphere coupled to a spectrometer and a charge-coupled device. The same solutions were spin casted on $Al_2O_3$-coated aluminum substrates and were mounted in the variable-temperature insert of a closed-cycle He cryostat with optical access for PL and RL measurements as a function of temperature. The PL spectra were excited using a frequency-tripled pulsed Nd:YAG laser source at 3.49 eV (355 nm) operated at 10 kHz; the emitted light was collected using a custom apparatus featuring a liquid-nitrogen-cooled, back-illuminated and UV-enhanced charge-coupled device detector (Jobin Yvon Symphony II) coupled to a monochromator (Jobin Yvon Triax 180) with 100 and 600 lines mm$^{-1}$ gratings. Time-resolved PL measurements were carried out using 3.06 eV (405 nm) picosecond pulsed laser diodes (Picoquant LDH-P series, ~70 ps pulses); the emitted light was collected with a phototube coupled to a Cornerstone 260 1/4 m visible–near-infrared monochromator (ORIEL) and a time-correlated single-photon counting unit (time resolution, ~400 ps).

*RL measurements*: CsPbBr$_3$ NCs spin casted onto Al$_2$O$_3$-coated substrates and nanocomposites were excited by unfiltered X-ray irradiation using a Philips PW2274 X-ray tube, with a tungsten target, equipped with a beryllium window and operated at 20 kV. At this operating voltage, a continuous X-ray spectrum is produced by a *Bremsstrahlung* mechanism superimposed to the L and M transition lines of tungsten, due to the impact of electrons generated through thermionic effect and accelerated onto a tungsten target. Cryogenic RL measurements are performed in the 10−290 K interval by lowering the temperature and using the same closed-cycle He cryostat used for the PL studies. The acquisition system is the same as the temperature-controlled PL measurements.

*Time-resolved radioluminescence*: The time response of scintillation light is measured in the microsecond timescale using a pulsed X-ray source composed by a 405 nm ps-pulsed laser (Edinburgh Instruments, EPL-405) hitting the photocathode of an X-ray tube from Hamamatsu (N5084) set at 40 kV. The emitted scintillation light was collected using a FLS980 spectrometer (Edinburgh Instruments) coupled to a PicoHarp 300 hybrid photomultiplier tube working in time-correlated single photon counting (TCSPC) mode (time resolution ∼150 ps).

*TSL and AG measurements:* wavelength-resolved TSL measurements at cryogenic temperatures are carried out by using the same detection system as for RL measurements on the same spin-casted CsPbBr$_3$ NCs and nanocomposites used for the RL experiments. Cryogenic TSL measurements are performed in the 10−300 K interval, with a heating rate of 0.1 K s$^{-1}$ after X-ray irradiation up to 5 Gy. The dose values for X-ray irradiation were obtained with a calibrated ionization chamber and evaluated in air. A typical TSL measurement consists in a preliminary X-ray irradiation at low temperature (T = 10 K in our case) to fill trap states in the bandgap that are stable at the chosen temperature: successively, the irradiation is ceased, and the temperature is progressively raised with a linear heating rate while monitoring the delayed emission due to carrier de-trapping. The shape of the TSL signal vs *T* (the so-called *glow curve*) has been corrected for the variation of the RL emission intensity vs *T*, to decouple the trap contribution to the emission from the other mechanisms involved in the scintillation. Wavelength-resolved AG measurements are carried out by monitoring the luminescence emission at a constant temperature (T = 10 K) as a function of delay time after the suppression of X-ray irradiation: the samples are irradiated with the same doses used for the TSL experiments.

*γ-ray irradiation experiments*: the same amount of NCs powders and four identical nanocomposite samples were placed in individual polypropylene sealed vials whose γ-ray attenuation is negligible (Eppendorf Tubes), one for each dose of NCs. The vials were irradiated in a pool-type γ-ray irradiation chamber equipped with a $^{60}$Co (mean energy, ~1.25 MeV) γ-source rod array, uniformly irradiating the NCs at 3.05 kGy$_{air}$h$^{-1}$ dose rate. CsPbBr$_3$ NCs and nanocomposites were irradiated at different cumulative absorbed doses up to 1 MGy by varying the irradiation time. The irradiation has been carried out at the gamma irradiation Calliope facility at ENEA-Casaccia Research Centre[2]. No γ-ray-attenuating material was interposed between the source and samples. Throughout the paper, the given dose is in air.

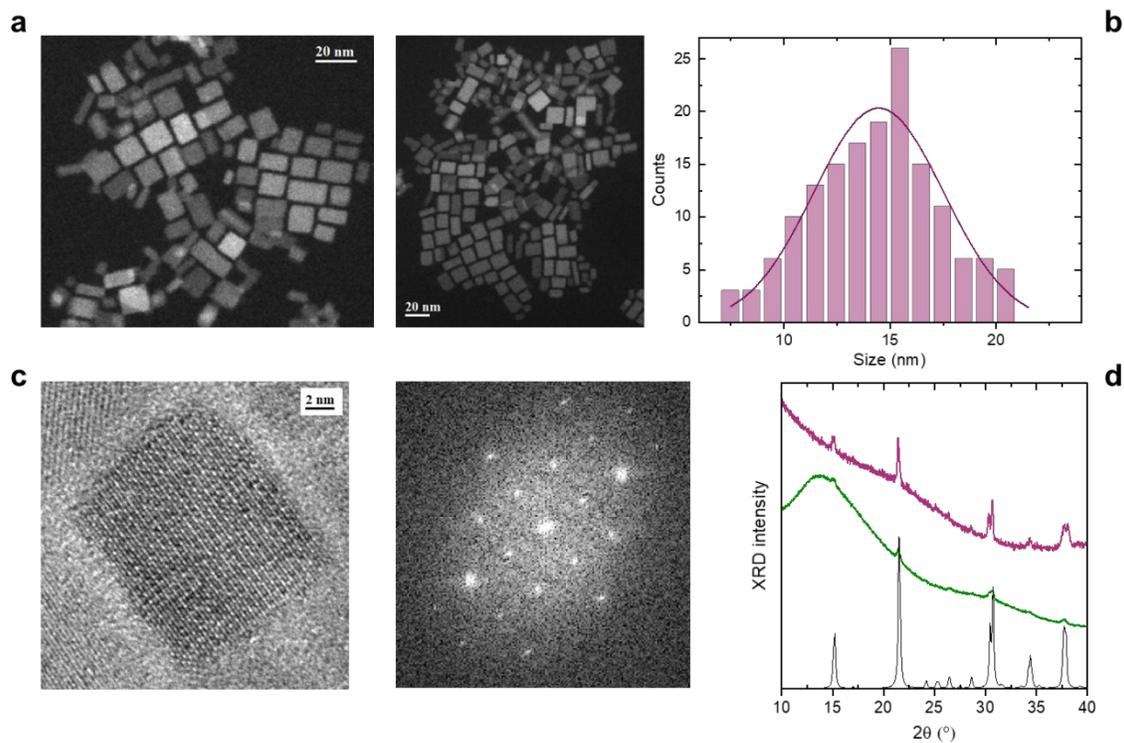

**Figure S1.** a) TEM images of CsPbBr$_3$ NCs. b) Size distribution of CsPbBr$_3$ NCs (14±3 nm) as extracted from TEM images. c) HRTEM of a CsPbBr$_3$ NC, with the corresponding Fast Fourier Transform. d) XRD patterns of CsPbBr$_3$ NCs (purple) and nanocomposite (green); the black curve corresponds to a calculated diffractogram for the orthorhombic phase of CsPbBr$_3$.

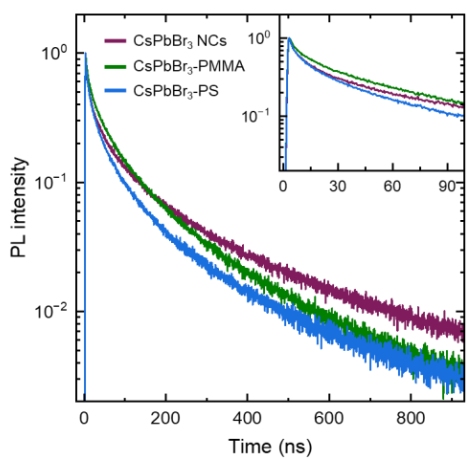

**Figure S2.** Normalized PL decay curves measured at the PL maximum by exciting with 3.05 eV light CsPbBr$_3$ NCs in toluene solution, after the embedding in PMMA and PS, as reported in the legend.

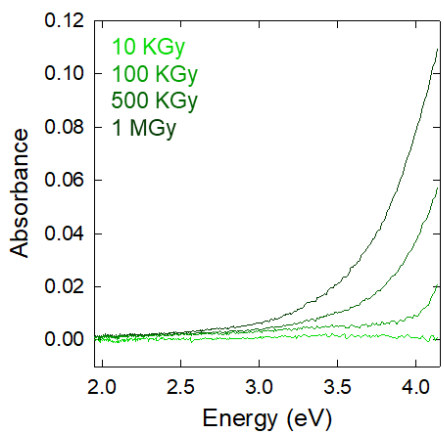

**Figure S3.** Optical absorption spectra of a film of bare PMMA as a function of cumulative irradiation dose up to 1 MGy from a $^{60}$Co source.

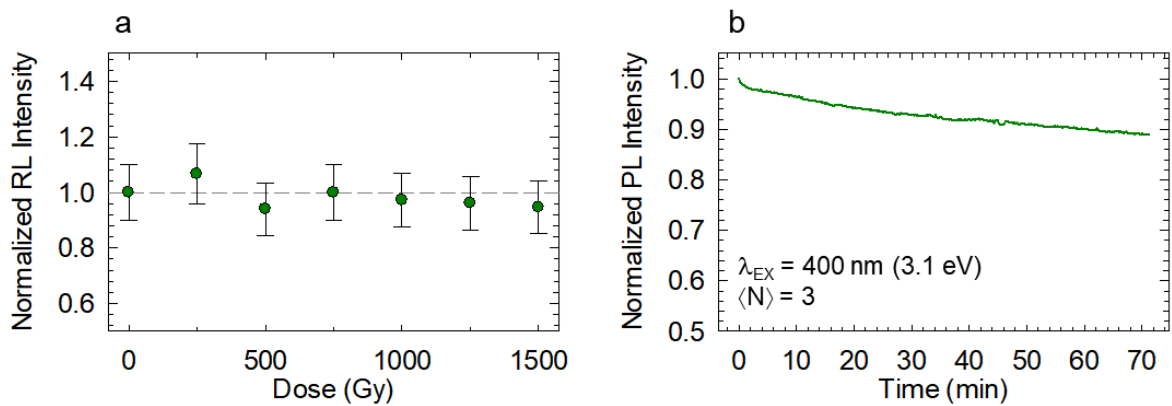

**Figure S4. a)** Normalized RL intensity of a PMMA nanocomposite as a function of cumulative irradiation dose of soft X-rays (E$_{max}$ = 25 kV) up to 1.5 kGy. The dashed line is a guide for the eye. **b)** Normalized PL intensity monitored over 70 minutes of continuous UV light excitation (E = 3.1 eV). The photon fluence was set to achieve an average excitonic population ⟨N⟩=3 corresponding to the excitation level obtained in scintillation measurements with 40 keV X-rays.

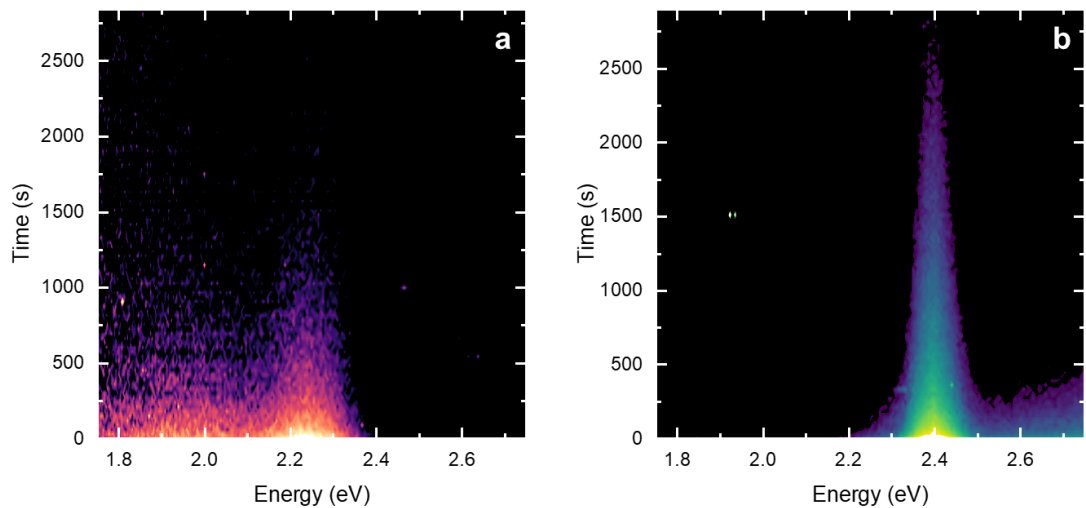

**Figure S5.** Contour plot of the spectrally resolved isothermal decay of AG intensity at 10 K as a function of time for CsPbBr$_3$ NCs **(a)** and nanocomposite **(b)**, obtained after X-ray irradiation up to ~ 5 Gy.

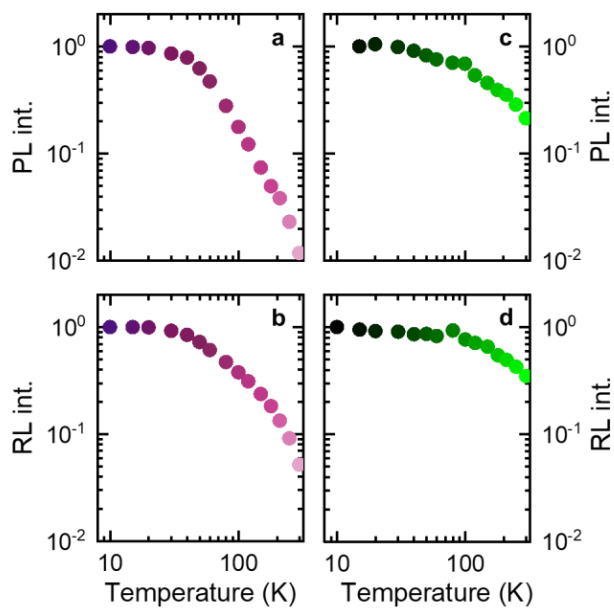

**Figure S6.** Spectrally integrated total PL (a) and RL (b) intensities as a function of temperature for low-QY CsPbBr$_3$ NCs. Spectrally integrated total PL (c) and RL (d) intensities as a function of temperature for the related nanocomposite.

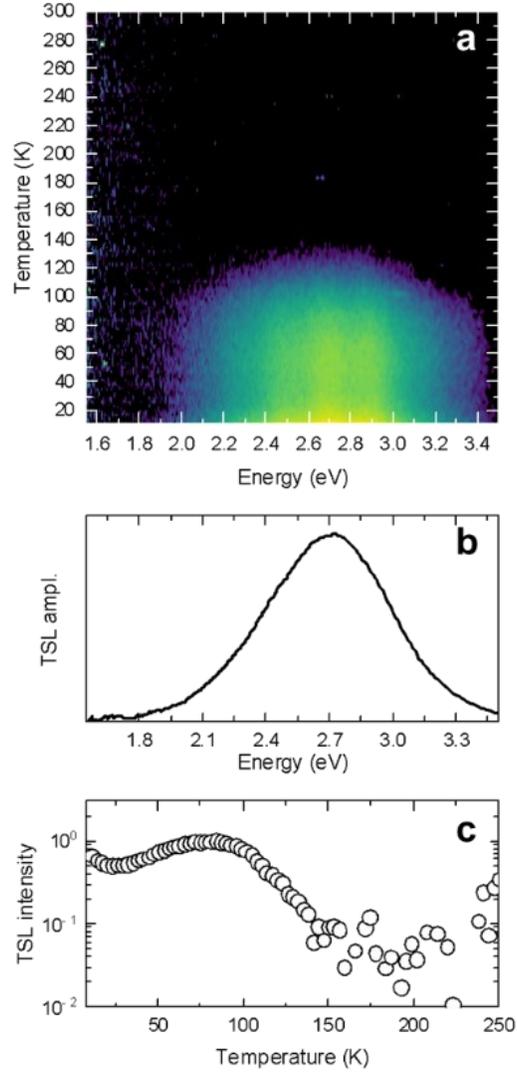

**Figure S7.** a) Contour plot of the spectrally resolved TSL as a function of temperature of a film of pure PMMA after X-ray irradiation at 10 K. b) Normalized TSL spectrum obtained by integrating the TSL emission in the 10 – 100 K range, and c) glow curve at 2.7 eV as extracted from (a).

|  | $\tau_F$ (ns) | $w_F$ (%) | $\tau_X$ (ns) | $w_X$ (%) | $\tau_S$ (ns) | $w_S$ (%) |
|---|---|---|---|---|---|---|
| **CsPbBr₃ NCs** | 6.8 | 7 | 77 | 43 | 455 | 50 |
| **CsPbBr₃ PMMA** | 10 | 12 | 94 | 63 | 455 | 25 |

**Table 1.** Fit parameters (decay time and relative weight, $w_i$) employed to analyze the time-resolved PL intensity decay spectra recorded on CsPbBr₃ NCs in solution and PMMA nanocomposite. The time-resolved PL discussed in the main text has been reproduced with triple exponential function.

$$I_{PL}(t) \propto \sum_{i=1}^{3} A_i e^{-(t/\tau_i)}$$

In the table, $\tau_F$ and $w_F$ are related to the fast component due to band-edge excitons affected by trapping, $\tau_X$ and $w_X$ to radiative exciton recombination, and $\tau_S$ and $w_S$ to delayed band-edge luminescence ascribed to thermal release of excitons trapped in shallow defect states.

|  | $w_P$ (%) | $\tau_1$ (ps) | $w_1$ (%) | $\tau_2$ (ps) | $w_2$ (%) |
| --- | --- | --- | --- | --- | --- |
| **CsPbBr$_3$ PMMA** | 5 | 675 | 29 | 7000 | 66 |

**Table 2.** Fit parameters (decay time and relative weight, $w_i$) employed to analyze the time-resolved RL intensity decay spectra recorded on CsPbBr$_3$ PMMA nanocomposite. The RL dynamics was analyzed in a least squares sense using the following formula, which also takes into account the convolution with the instrument response function (IRF). Specifically, the decays were modelled analytically using two exponential decay functions summed to a prompt component, here represented by a Gaussian pulse much shorter than the IRF temporal dispersion.

$$F(t) = IRF(t) \otimes \left( a_G \cdot e^{-\frac{(t-t_0)^2}{2\sigma^2}} + H(t-t_0) \cdot \left[ \sum_{i=1}^{2} a_i \cdot e^{-t/\tau_i} \right] \right) + C$$

In the equation above, $t_0$ corresponds to the start of the emission process, $C$ is the electronic background noise floor, and $H$ is the Heaviside function. The experimental IRF was well reproduced by a Gaussian profile (FWHM = 150 ps) and the weight of each component (prompt and single exponential decays) is evaluated as the integral of each convoluted function over the entire time window.

According to ref [3], the average lifetime was calculated with the re-normalized ratio of components $\tau_1$ and $\tau_2$ according to $<\tau> = (w_{1n}/\tau_1 + w_{2n}/\tau_2)^{-1}, w_{in} = w_i/(w_1 + w_2)$.